\documentclass{jpsj-suppl}
\usepackage{times}
\usepackage{color}
\usepackage{bm}
\usepackage{amsmath}

\recdate{September 30, 2013}

\title{Impurity Effect on the Local Density of States around a Vortex in Noncentrosymmetric Superconductors}

\author{
Yoichi \textsc{Higashi}\thanks{E-mail address: higashiyoichi@ms.osakafu-u.ac.jp}$^{,1,3}$,
Yuki \textsc{Nagai}$^{2}$
, and Nobuhiko \textsc{Hayashi}$^{3}$
}

\inst{
$^{1}$Department of Mathematical Sciences, Osaka Prefecture University, 1-1 Gakuen-cho, Naka-ku, Sakai 599-8531, Japan\\
$^{2}$CCSE, Japan Atomic Energy Agency, 5-1-5 Kashiwanoha, Kashiwa, Chiba 277-8587, Japan\\
$^{3}$Nanoscience and Nanotechnology Research Center (N2RC), Osaka Prefecture University, 1-2 Gakuen-cho, Naka-ku, Sakai 599-8570, Japan
}

\abst{
We numerically study the effect of non-magnetic impurities on the vortex bound states in noncentrosymmetric systems.
The local density of states (LDOS) around a vortex is calculated
by means of the quasiclassical Green's function method.
We find that the zero energy peak of the LDOS splits off with increasing the impurity scattering rate.
}

\kword{Noncentrosymmetric sysyem, Vortex core, Local density of states, Impurity effect}

\begin{document}
\maketitle
\section{Introduction}
Much attention has been focused on the superconductivity in noncentrosymmetric systems
because of its novel superconductivity due to spin-orbit coupling (SOC).
Recently,
active investigations have been conducted
on the inhomogenious superconducting state in noncentrosymmetric systems
such as the helical phase in a magnetic field \cite{kaur2005}, the vortex state \cite{yip2005,hayashi2006},
and the exotic superconducting state in locally noncentrosymmetric systems \cite{yoshida2012}.

In this study, we investigate the non-magnetic impurity scattering effect on the local density of states (LDOS) around a single vortex.
Impurity effects on the vortex core structure in noncentrosymmetric systems can differ from the simple two-gap systems
because there is SOC in this system.
So unusual phenomena are expected in the LDOS inside a vortex core.
There are several previous studies for the impurity effect on the noncentrosymmetric superconductivity
in the bulk without vortices \cite{frigeri2006,mineev2007}.
However,
there is no research on the impurity effect
in spatially inhomogeneous situations such as vortex state
in noncentrosymmetric systems.

Through our formulation using the quasiclassical Green's function method,
it turned out that
the form of the impurity self energy around a vortex is quite different from that in the bulk.
This comes from the effect of the superflow on the impurity self energy.
In addition,
the Green's function cannot be separated
with respect to each Fermi surface (FS) split due to the Rashba-type SOC
in the presence of both impurities and vortices.
This fact might also influence the impurity effect on the vortex core structure.

In this paper,
we numerically investigated the effect of non-magnetic impurities on the LDOS around a vortex.
We found that the zero-energy peak (ZEP) of the LDOS splits off with increasing the impurity scattering rate.
We give the rough physical interpretation for the numerical results.
\section{Formulation}
The noncentrosymmetricity of the system induces the anitisymmetric SOC
and permits the mixing of the Cooper pair wave function with different parity.
We assume the Rashba-type SOC, which is described as $\bm{g}_k=\sqrt{3/2}(-{k}_y,{k}_x,0)/k_{\rm F}$
by the orbital vector $\bm{g}_k$.
$k_{\rm F}$ is the Fermi wave number.
We consider a single vortex line along the $z$ axis situated at the origin ($\bm{r}=0$).
We assume that the FS is spherical.
The lack of spatial inversion symmetry of the system is detrimental to spin-triplet state \cite{anderson1984}.
However,
the spin-triplet state is not suppressed
under the particular situation in which $\bm{d}_k~||~\bm{g}_k$ \cite{frigeri2004}.
Then we consider the parity mixing of the Cooper pair.

The parity-mixed pairing state is expressed as
$
\hat{\varDelta}(\bm{r},\tilde{\bm{k}})=\left[ \varPsi(\bm{r})\hat{\sigma}_0 + \bm{d}_k(\bm{r}) \cdot \hat{\bm{\sigma}} \right]i\hat{\sigma}_y
=[ \varPsi \hat{\sigma}_0 + \varDelta ( -\tilde{k}_y\hat{\sigma}_x + \tilde{k}_x \hat{\sigma}_y ) ] f(r) \exp[i\phi_r] i\hat{\sigma}_y,
$
where the spin-singlet $s$-wave component $\varPsi(\bm{r})$,
the $d$ vector $\bm{d}_k=\varDelta(\bm{r})(-\tilde{k}_y,\tilde{k}_x,0)$
with the unit vector $\tilde{\bm{k}}=(\tilde{k}_x,\tilde{k}_y,\tilde{k}_z)=(\cos \phi_k \sin \theta_k, \sin \phi_k \sin \theta_k, \cos \theta_k)$,
$\varPsi (\varDelta)$ is the bulk amplitude of the pair potential for singlet (triplet) component,
$\hat{\bm{\sigma}}=(\hat{\sigma}_x,\hat{\sigma}_y,\hat{\sigma}_z)$ is the vector consisted of the Pauli spin matrices
and $\hat{\sigma}_0$ is the unit matrix in the spin space.
We assume that the both components of the pair potential have the same real space profile around a vortex
and set $f(r)=r/\sqrt{r^2+\xi^2_0}$.
Here, $\xi_0=v_{\rm F}/T_{\rm c}$ is the coherence length,
$T_{\rm c}$ is the superconducting critical temperature,
$v_{\rm F}(=|\bm{v}_{\rm F}|)$ is the Fermi velocity.
In this paper,
we does not solve the gap equations for the pair potentials $\varPsi(\bm{r})$ and $\varDelta(\bm{r})$ self-consistently.
We conduct the numerical calculation for zero temperature and 
take into account the effect of temperature through the energy smearing factor $\eta$.

In order to obtain the LDOS and the impurity self energy,
we consider the following quasiclassical Green's function, which is defined in the particle-hole space.
\begin{equation}
\check{g}(\bm{r},\tilde{\bm{k}},i\omega_n)=-i\pi
\left(
\begin{array}{cc}
\hat{g} & i\hat{f} \\
-i\hat{\bar{f}} & -\hat{\bar{g}}
\end{array}
\right).
\end{equation}
Here, $\omega_n$ is the Matsubara frequency.
$\hat{\cdot}$ denotes the $2 \times 2$ matrix in the spin space and
$\check{\cdot}$ denotes the $4 \times 4$ matrix in the particle-hole and spin space.
The Eilenberger equation with the SOC term and impurity self energy is given as \cite{serene1983}
\begin{equation}
i\bm{v}_{\rm F}(\tilde{\bm{k}}) \cdot \bm{\nabla} \check{g}(\bm{r},\tilde{\bm{k}},i\omega_n)
+\left[ i\omega_n \check{\tau}_3 -\check{\varDelta} -\check{\varSigma} -\alpha \check{\bm{g}}_k \cdot \check{\bm{S}},~\check{g}(\bm{r},\tilde{\bm{k}},i\omega_n) \right]=\check{0},
\label{Eilenberger eq.}
\end{equation}
where
\begin{alignat}{1}
\check{\tau}_3
  &=
  \left(
\begin{array}{cc}
\hat{\sigma}_0 & \hat{0} \\
\hat{0} & -\hat{\sigma}_0
\end{array}
\right),
~~~
\check{\bm{S}}
    =
    \left(
\begin{array}{cc}
\hat{\bm{\sigma}} & \hat{0} \\
\hat{0} & \hat{\bm{\sigma}}^{\rm tr}
\end{array}
\right),
~~~
\hat{\bm{\sigma}}^{\rm tr}
=
-\hat{\sigma}_y \hat{\bm{\sigma}} \hat{\sigma}_y,
\\
  \check{\bm{g}}_k
  &=
  \left(
\begin{array}{cc}
\bm{g}_k\hat{\sigma}_0 & \hat{0} \\
\hat{0} & \bm{g}_{-k}\hat{\sigma}_0
\end{array}
\right),
~~~
\check{\varDelta}=
\left(
\begin{array}{cc}
\hat{0} & \hat{\varDelta}(\bm{r},\tilde{\bm{k}}) \\
-\hat{\varDelta}^\dag(\bm{r},\tilde{\bm{k}}) & \hat{0}
\end{array}
\right).
\end{alignat}
Here, $\alpha$ is the strength of the SOC.
$\bm{g}_k=\sqrt{3/2}(-\tilde{k}_y,\tilde{k}_x,0)$,
which is odd in $k$ (i.e., $\bm{g}_{-k}=-\bm{g}_k$).
We use the units in which $\hbar=k_{\rm B}=1$.
The quasiclassical Green's function satisfies the normalization condition $\check{g}^2=-\pi^2 \check{\tau}_0$
with the $4 \times 4$ unit matrix $\check{\tau}_0$.
The bracket $[\cdots,\cdots]$ is a commutator.

The Eilenberger equation is separated into two equations
without spin degree of freedom
using the band basis both in the clean vortex state \cite{hayashi2006} and in the bulk clean system \cite{hayashiPRB2006,hayashi2006nmr}.
The normal-state Hamiltonian in the clean limit, which is a $2\times2$ matrix in the spin space, becomes diagonal in the band basis.
The off-diagonal components of the quasiclassical Green's function decay to zero around a vortex
and the quasiclassical Green's function in the band basis becomes diagonal in the spin space (see the appendix of Ref.~\cite{hayashiPRB2006}).
Each diagonal component obeys the Eilenberger equation, which is defined on the two FSs split due to the SOC.
Then,
we can solve the Eilenberger equation without the SOC term
with respect to each FS
in the clean limit.
This situation is not changed in the presence of impurities in the bulk.

However,
the quasiclassical Green's function is not diagonal in the band basis
in the presence of both impurities and vortices.
Therefore the Eilenberger equation is not separated into two equations with respect to each FS.
Thus, we use the orbital basis, in which the spin quantization axis is oriented parallel to the $z$ axis.
Under this basis,
we transform Eq.~\eqref{Eilenberger eq.} into the two matrix Riccati type differential equations,
which is the first-order differential equations [see Appendix].
Then we obtain the stable numerical solution
solving the Eq.~\eqref{riccati_a} and \eqref{riccati_b} by means of the adaptive stepsize control Runge-Kutta method \cite{numerical1992}.

In this paper,
we investigate the impurity scattering in the Born limit (We set the scattering phase shift exactly zero).
In this limit,
the impurity self energy is given as \cite{kato2000,hayashi2005}
\begin{align}
\check{\varSigma}(\bm{r},E)
 &= \cfrac{\varGamma_{\rm n}}{\pi}\left\langle \check{g}_0(\bm{r},\tilde{\bm{k}},i\omega_n \rightarrow E+i\eta)  \right\rangle_{\tilde{\bm{k}}},
 \\
 &= \varGamma_{\rm n}
 \left(
\begin{array}{cc}
-i\langle \hat{g} \rangle_{\tilde{\bm{k}}} & \langle \hat{f} \rangle_{\tilde{\bm{k}}}  \\
-\langle \hat{\bar{f}} \rangle_{\tilde{\bm{k}}} & i\langle \hat{\bar{g}} \rangle_{\tilde{\bm{k}}}
\end{array}
\right),
\end{align}
where $\varGamma_{\rm n}$ is the impurity scattering rate in the normal state,
$\check{g}_0$ is the quasiclassical Green's function in the clean limit,
$\left \langle \cdots \right \rangle_{\tilde{\bm{k}}}$ denotes the average over the FS
and $\check{\varSigma}\equiv\{ \hat{\varSigma}_{ij} \}_{i,j=1,2}$.
Now we consider the system with the rotational symmetry about a vortex line.
In the clean limit,
the parity mixing pair potential has a form
$\varDelta_{\rm I,II}(r,\phi_r;\theta_k)=\left[ \varPsi(r) \pm \varDelta(r) \sin \theta_k \right] \exp\left[ i\phi_r \right]$
around a vortex using the band basis \cite{hayashi2006}.
These forms of the pair potentials have the $\theta_k$ dependence only.
We assume that the pairing state does not change with non-magnetic impurities.
Under an axial rotation about a vortex line,
the pair potential has the azimuthal angle dependence
in the form of the phase factor in the real space.
From Eq.~\eqref{riccati_a} and \eqref{riccati_b},
we can see that the anomalous self energies $\hat{\varSigma}_{12,21}$
have the same phase factor as the pair potentials,
whereas the normal self energies $\hat{\varSigma}_{11,22}$ are invariant under the axial rotation.
Thus the anomalous self energies $\hat{\varSigma}_{12,21}$ have the following $\phi_r$ dependence \cite{hayashi2005}:
$
\hat{\varSigma}_{12,21}(r,\phi_r,E)  = \hat{\varSigma}_{12,21}(r,E) \exp[\pm i\phi_r].
$
The plus (minus) sign corresponds to $\hat{\varSigma}_{12} (\hat{\varSigma}_{21})$.
In the actual numerical calculation,
the self energies are discretely calculated in a radial $r$ direction.
Thus they are linearly-interpolated in this direction.
As for the direction of an azimuthal angle,
the self energies have the continuous $\phi_r$ dependence.

The LDOS per spin is obtained from
\begin{equation}
N(\bm{r},E)=-\cfrac{N_{\rm F}}{2}\cfrac{1}{\pi}\left\langle {\rm Im}\left[ {\rm Tr}~\hat{g}(\bm{r},\tilde{\bm{k}},i\omega_n \rightarrow E+i\eta) \right] \right\rangle_{\tilde{\bm{k}}},
\end{equation}
where $N_{\rm F}$ is the density of states per spin at the Fermi level in the normal state.

\section{Result and Discussion}
\begin{figure}[tb]
\begin{center}
    \begin{tabular}{p{70mm}p{80mm}}
      \resizebox{70mm}{!}{\includegraphics{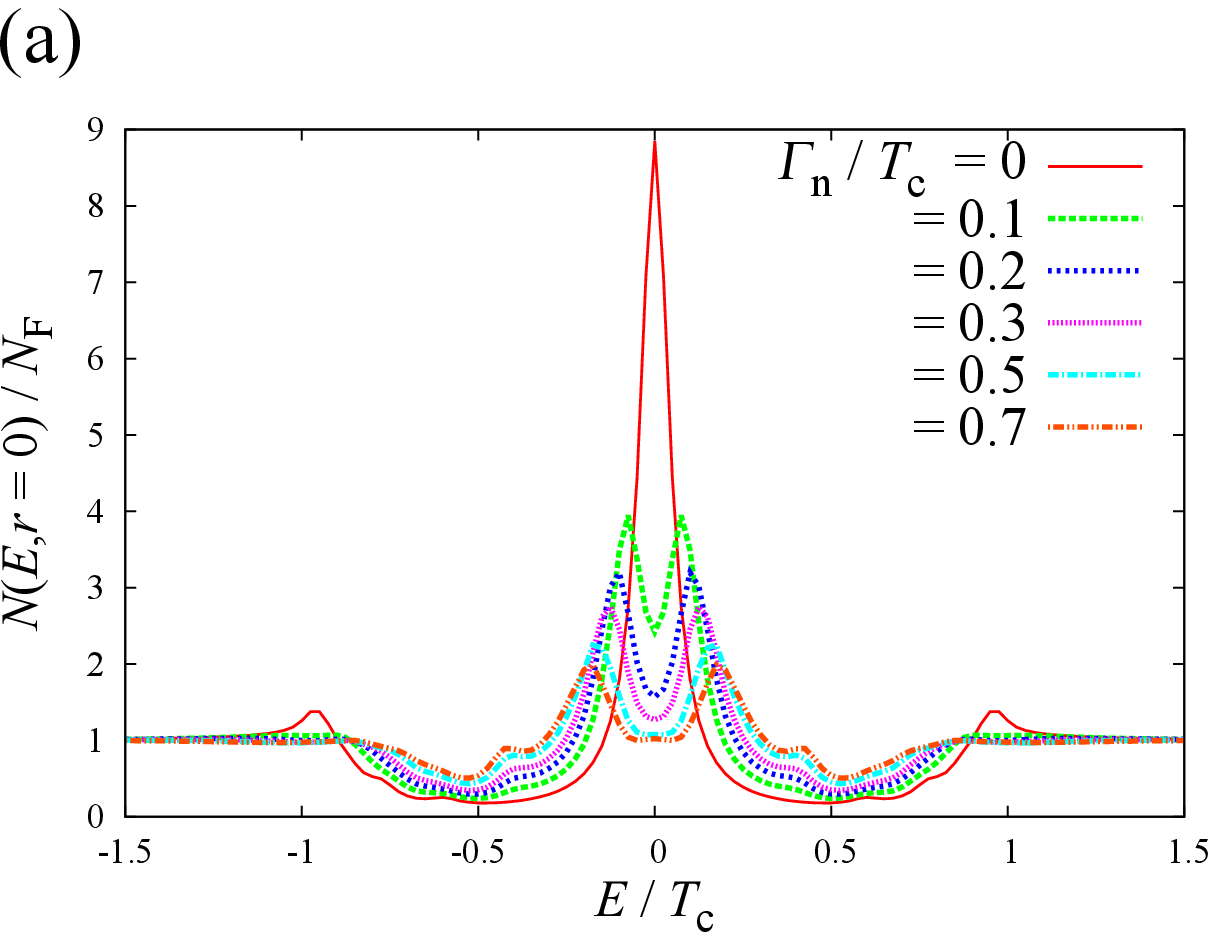}} &
      \resizebox{80mm}{!}{\includegraphics{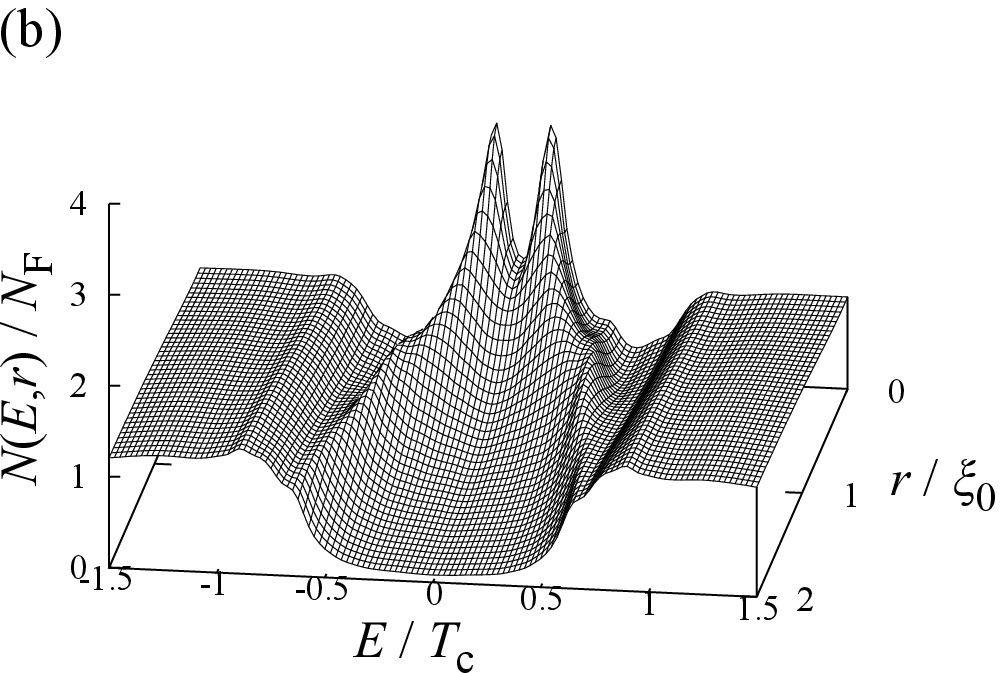}}
    \end{tabular}
\caption{
(Color online)
$s$-wave case.
The energy smearing factor $\eta$ is set to $0.05T_{\rm c}$.
(a) The energy dependence of the local density of states at the vortex center for several values of the impurity scattering rate $\varGamma_{\rm n}$.
(b) The local density of states within a vortex core for $\varGamma_{\rm n}=0.3T_{\rm c}$.
}
\label{fig1}
\end{center}
\end{figure}

Throughout this section, we set $\alpha^\prime/T_{\rm c}=1$
with $\alpha^\prime=\sqrt{3/2}\alpha$.
We show in Fig.~\ref{fig1}(a)
the energy dependence of the LDOS at the vortex center
for several values of the impurity scattering rate $\varGamma_{\rm n}$
for the $s$-wave case ($\varPsi/T_{\rm c}=1,\varDelta/T_{\rm c}=0$).
$\varGamma_{\rm n}/T_{\rm c}$ is roughly estimated of the order of $\xi_0/l$.
$l$ is the mean free path.
In the clean limit ($\varGamma_{\rm n}/T_{\rm c}=0$),
the peak of the LDOS is seen at the zero energy.
In the presence of impurities,
the ZEP splits into two peaks.
We confirmed that the split of the ZEP does not occur for $\alpha^\prime/T_{\rm c}=0$.
The splitting width becomes larger and the peak height decreases
with increasing the impurity scattering rate $\varGamma_{\rm n}$.
For $\varGamma_{\rm n}=0.7T_{\rm c}$,
the value of the zero energy LDOS at the vortex center
equals to the normal state density of states $N_{\rm F}$.
In Fig.~\ref{fig1}(b),
we show the energy and the spatial profile of the LDOS inside a vortex core for $\varGamma_{\rm n}=0.3T_{\rm c}$.
We see clearly that the ZEP of the LDOS splits off at the vortex center.
Far from the vortex center,
the zero energy density of states decays to zero and
the fully opened superconducting energy gap appears
in the bulk,
which is understood from the Anderson's theorem for non-magnetic impurities \cite{anderson1959}.
We can see that the ridge lines of the vortex bound states
approach the gap edges of the $s$-wave component $E/T_{\rm c}=\varPsi/T_{\rm c}=\pm1$.

\begin{figure}[tb]
  \begin{center}
    \begin{tabular}{p{70mm}p{80mm}}
      \resizebox{70mm}{!}{\includegraphics{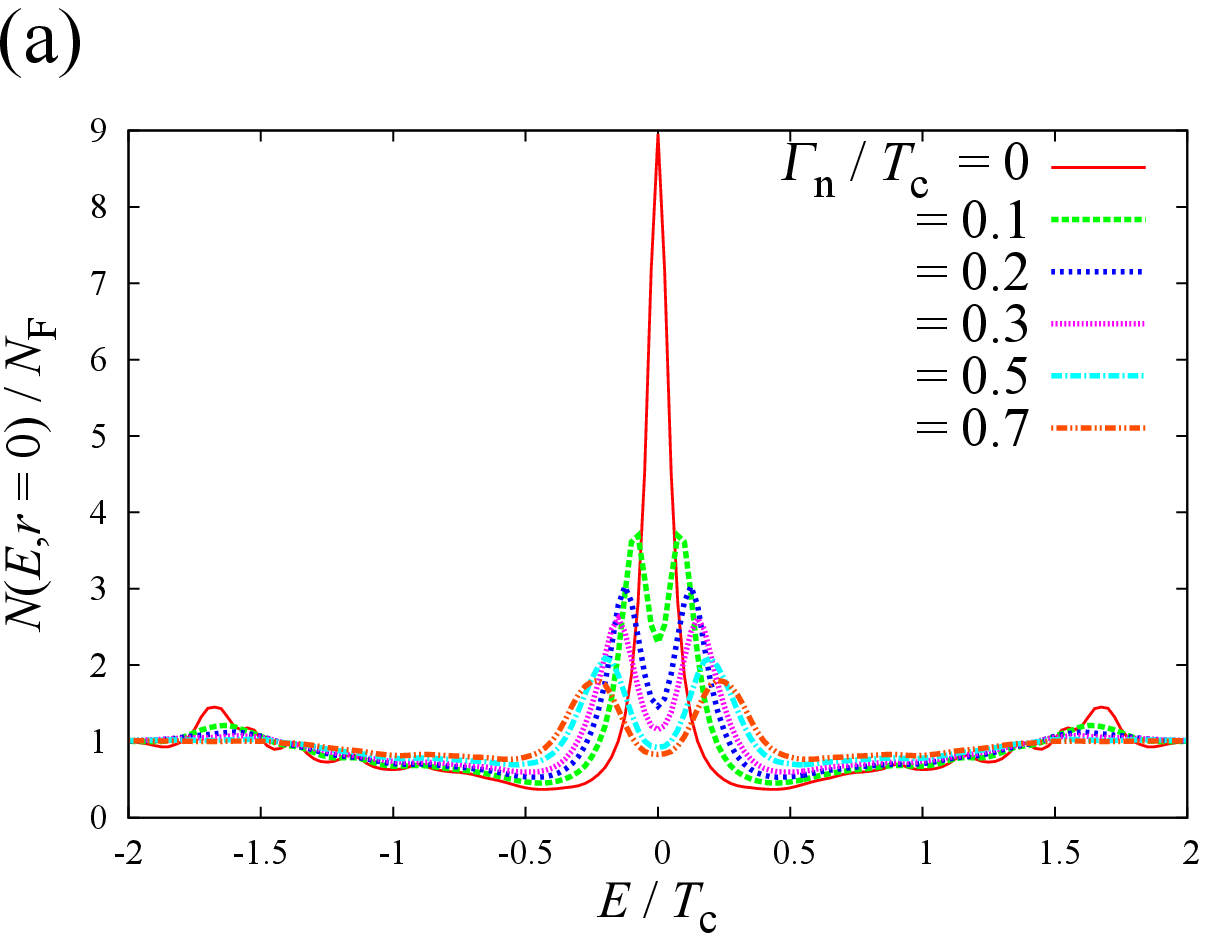}} &
      \resizebox{80mm}{!}{\includegraphics{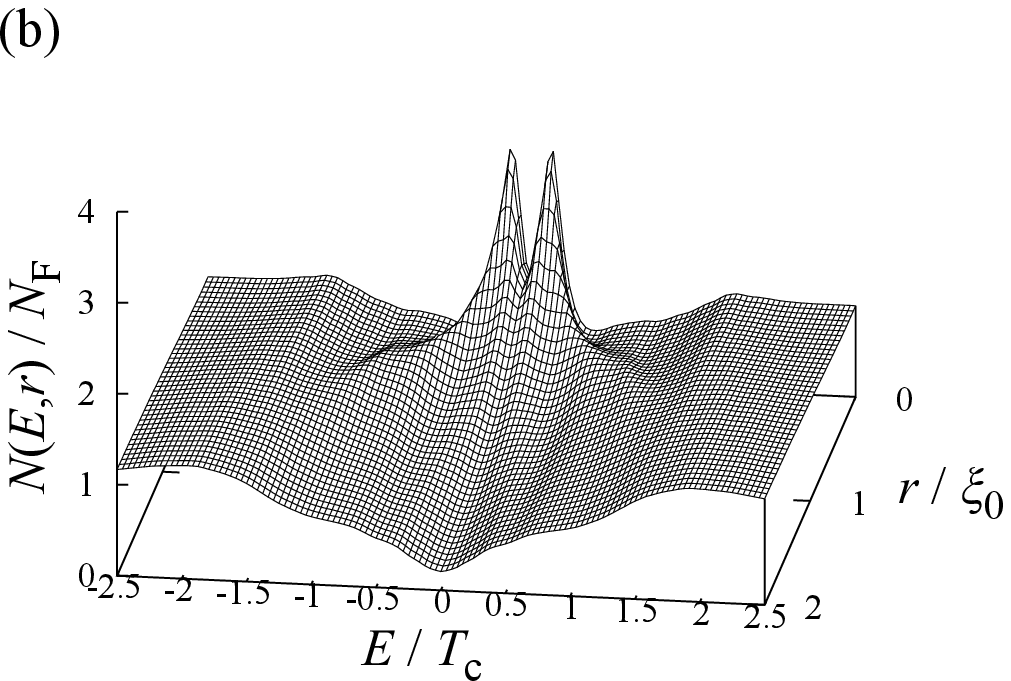}}
    \end{tabular}
\caption{
(Color online)
$s+p$-wave case.
The energy smearing factor $\eta$ is set to $0.05T_{\rm c}$.
(a) The energy dependence of the local density of states at the vortex center for several values of the impurity scattering rate $\varGamma_{\rm n}$.
(b) The local density of states within a vortex core for $\varGamma_{\rm n}=0.3T_{\rm c}$.
}
\label{fig2}
\end{center}
\end{figure}

As shown in Fig.~\ref{fig2}(a),
the ZEP splits off
also in the case of the $s+p$-wave ($\varPsi/T_{\rm c}:\varDelta/T_{\rm c}=0.7:1.4$).
The ridge lines related to the vortex bound states of the quasiparticles
on the split FS are broaden due to impurity scattering [see Fig.~\ref{fig2}(b)].
We cannot observe the two-gap like quasiparticle spectra within a core,
which is seen in the clean limit \cite{hayashi2006}.
Within the superconducting gap energy,
there exists the non-zero density of states
which comes from excitations in the vicinity of the horizontal line nodes.
Nodes can appear under $\varPsi < \varDelta$ \cite{hayashi2006nmr},
which corresponds to the present condition.


Let us explain about the present physical interpretation of the split ZEP.
The condition that the Green's functions are invariant under $\bm{k}\rightarrow -\bm{k}$
is satisfied at least in spatially uniform systems,
when $\varDelta_{\rm I,II}$ are invariant under $\bm{k}\rightarrow -\bm{k}$.
In this situation,
$\hat{\varSigma}^0_{ij} \propto \hat{\sigma}_0$.
Therefore,
the impurity scattering effect
appears only in the modification of the Matsubara frequency and the pair potential in the band basis.
The Green's function is still diagonal in the band basis.
That is,
the split two bands are not mixed by impurity scattering in the bulk
within the Born approximation.
Frigeri {\it et al.,} have already pointed out this fact \cite{frigeri2006}.

However,
the condition is not satisfied in systems where the supercurrent flows (e.g., around a vortex core),
namely the Green's functions are
not invariant under $\bm{k}\rightarrow -\bm{k}$ in such a system.
So the situation at the vortex core is quite different from that in the bulk.
Through a careful examination,
it turns out that
the anomalous self energies $\hat{\varSigma}^0_{12}$ and $\hat{\varSigma}^0_{21}$ are not diagonalized
in the band basis.
They contribute to mix the two bands split due to the SOC.
If regarding them as the perturbation
by which the transition occurs
between the bound state spectra of quasiparticles on the each split FS,
it might be considered that
the overlap or the crossing of the vortex bound state spectra is resolved to give a energy gap.
This situation is similar to
the two bands separated by a band gap at the corners of the Brillouin-zone.


\section{Conclusion}
In conclusion,
we calculated the local density of states around a vortex core
in the presence of non-magnetc impurities
in noncentrosymmetric systems.
We found that the zero-energy peak of the vortex bound states splits off
with increasing the impurity scattering rate
for both $s$-wave and $s+p$-wave pairing state.
In the noncentrosymmetric systems,
the impurity effect on vortex bound states exhibits the different one
comparing with that of simple two-gap superconducting systems.
The spin-orbit coupling is essential for this phenomena.

\section{Acknowledgments}
We thank Y. Masaki for discussions from the more microscopic point of view
and also thank N. Nakai and M. Kato for helpful discussions.

\appendix
\section{Matrix Riccati equation}
In this appendix,
we describe briefly the derivation of the matrix Riccati equation
including the Rashba-type spin-orbit coupling term and the impurity self energy
by means of the projection method \cite{eschrig2000}.
We introduce the projecters:
\begin{equation}
\check{P}_\pm=\cfrac{1}{2}\left( \check{\tau}_0 \pm \cfrac{1}{-i\pi}\check{g} \right).
\end{equation}
These projectors satisfy the relations:
\begin{alignat}{2}
\check{P}_\pm \cdot \check{P}_\pm  &= \check{P}_\pm,
\label{eq:A2}
\\
\check{P}_+ + \check{P}_-  &= \check{\tau}_0,
\label{closure}
\\
\check{P}_+ \cdot \check{P}_-  &= \check{P}_- \cdot \check{P}_+ &= \check{0}.
\label{orthogonal}
\end{alignat}
We can confirm the following relation: $\check{P}_+ - \check{P}_- = \check{g}/(-i\pi)$.
Then, the quasiclassical Green's function is obtained as $\check{g}=-i\pi\left( \check{P}_+ - \check{P}_- \right)$.
Substituting this Green's function into Eq.~\eqref{Eilenberger eq.}, we have the equation for the projectors $\check{P}_\pm$:
\begin{equation}
-i\bm{v}_{\rm F} \cdot \bm{\nabla} \check{P}_\pm = \left[ i\omega_n \check{\tau}_3 -\check{\varDelta} -\check{\varSigma} - \alpha \check{\bm{g}}_k \cdot \check{S},\check{P}_\pm \right].
\label{eq:P}
\end{equation}
We define the projectors as
\begin{alignat}{1}
\check{P}_+  &=
 \left(
\begin{array}{c}
\hat{\sigma}_0  \\
 -i\hat{b}
\end{array}
\right)
\left(
\hat{\sigma}_0 + \hat{a} \hat{b}
\right)^{-1}
\left(
\begin{array}{cc}
\hat{\sigma}_0 & i\hat{a}
\end{array}
\right),
\label{P_+}
\\
\check{P}_-  &=
\left(
\begin{array}{c}
-i\hat{a}  \\
 \hat{\sigma}_0
\end{array}
\right)
\left(
\hat{\sigma}_0+\hat{b}\hat{a}
\right)^{-1}
\left(
\begin{array}{cc}
i\hat{b} & \hat{\sigma}_0
\end{array}
\right),
\label{P_-}
\end{alignat}
which satisfy Eq.~\eqref{eq:A2} and \eqref{orthogonal}.
To satisfy Eq.~\eqref{closure},
we need the following relations between $\hat{a}$ and $\hat{b}$:
$
( \hat{\sigma}_0 +\hat{a}\hat{b} )^{-1} \hat{a}  = \hat{a} ( \hat{\sigma}_0 + \hat{b} \hat{a} )^{-1},~
\hat{b}( \hat{\sigma}_0 + \hat{a} \hat{b} )^{-1} = ( \hat{\sigma}_0 + \hat{b} \hat{a} )^{-1} \hat{b}.
$
Using these relations, the quasiclassical Green's function is obtained as
\begin{equation}
\check{g}=-i\pi
\left(
\begin{array}{cc}
\left( \hat{\sigma}_0 +\hat{a} \hat{b} \right)^{-1} \left( \hat{\sigma}_0 -\hat{a} \hat{b}\right) & 2i \left( \hat{\sigma}_0 +\hat{a}\hat{b} \right)^{-1} \hat{a} \\
-2i\hat{b} \left( \hat{\sigma}_0 +\hat{a}\hat{b} \right)^{-1} & -\left( \hat{\sigma}_0 + \hat{b} \hat{a} \right)^{-1} \left( \hat{\sigma}_0 -\hat{b} \hat{a} \right)
\end{array}
\right).
\end{equation}
Substituting Eq.~\eqref{P_+} and \eqref{P_-} into Eq.~\eqref{eq:P} and calculating with respect to each component,
we obtaine the matrix Riccati equations:
\begin{align}
\bm{v}_{\rm F} \cdot \bm{\nabla}\hat{a}_0 + 2\omega_n \hat{a}_0 & +\hat{a}_0\left(\hat{\varDelta}^\dag_0 -\hat{\varSigma}^0_{21} \right) \hat{a}_0 -\left( \hat{\varDelta}_0 + \hat{\varSigma}^0_{12} \right)
\nonumber \\
&
+i\alpha \bm{g}_k \cdot \left( \hat{\bm{\sigma}}\hat{a}_0-\hat{a}_0 \hat{\bm{\sigma}} \right) +i\left( \hat{\varSigma}^0_{11}\hat{a}_0+\hat{a}_0\hat{\varSigma}^0_{22} \right)= \hat{0},
\label{riccati_a}
\\
\bm{v}_{\rm F} \cdot \bm{\nabla}\hat{b}_0 - 2\omega_n \hat{b}_0 & -\hat{b}_0 \left( \hat{\varDelta}_0 + \hat{\varSigma}^0_{12} \right) \hat{b}_0 + \left( \hat{\varDelta}^\dag_0 - \hat{\varSigma}^0_{21}\right)
\nonumber \\
&
+i\alpha \bm{g}_k \cdot \left( \hat{b}_0 \hat{\bm{\sigma}} - \hat{\bm{\sigma}} \hat{b}_0 \right)-i \left( \hat{b}_0\hat{\varSigma}^0_{11} + \hat{\varSigma}^0_{22} \hat{b}_0 \right) = \hat{0}.
\label{riccati_b}
\end{align}
Here, we define 
$
\hat{a}=\hat{a}_0i\hat{\sigma}_y,
\hat{b}=-i\hat{\sigma}_y\hat{b}_0,
\hat{\varSigma}_{11}=\hat{\varSigma}_{11}^0,
\hat{\varSigma}_{12}=\hat{\varSigma}_{12}^0i\hat{\sigma}_y,
\hat{\varSigma}_{21}=-i\hat{\sigma}_y\hat{\varSigma}_{21}^0,
\hat{\varSigma}_{22}=-\hat{\sigma}_y \hat{\varSigma}^0_{22}\hat{\sigma}_y,
\hat{\varDelta}=\hat{\varDelta}_0 i\sigma_y,
\hat{\varDelta}^\dag=-i\sigma_y \hat{\varDelta}_0^\dag
$.

\end{document}